\documentclass[times,twocolumn,review]{elsarticle}
\usepackage{medima}
\usepackage{framed,multirow}
\usepackage{subfigure}
\usepackage{graphicx}
\usepackage{textcomp}
\usepackage{blindtext}

\usepackage[namelimits]{amsmath} %数学公式
\usepackage{amssymb}             %数学公式
\usepackage{amsfonts}            %数学字体
\usepackage{mathrsfs}            %数学花体
\usepackage{latexsym}
\usepackage{booktabs}
\usepackage{lineno}
\modulolinenumbers[5]
\usepackage{lineno}

\usepackage{url} % arXiv
\usepackage[table,xcdraw]{xcolor}
\usepackage{hyperref}  % arXiv

\begin{document}

\verso{Gongning Luo \textit{et~al.}}

\begin{frontmatter}

% \title{Efficient automatic segmentation for multi-level pulmonary arteries: The PARSE challenge\tnoteref{tnote1}}%
% \tnotetext[tnote1]{This is an example for title footnote coding.}
\title{Efficient automatic segmentation for multi-level pulmonary arteries: The PARSE challenge}%
% \tnotetext[tnote1]{This is an example for title footnote coding.}

% \author[1]{Given-name1 {Surname1}\corref{cor1}}
% \cortext[cor1]{Corresponding author: 
%   Tel.: +0-000-000-0000;  
%   fax: +0-000-000-0000;}
% \author[1]{Given-name2 {Surname2}\fnref{fn1}}
% \fntext[fn1]{This is author footnote for second author.}
% \author[2]{Given-name3 {Surname3}}
% %% Third author's email
% \ead{author3@author.com}
% \author[2]{Given-name4 {Surname4}}

% \address[1]{Affiliation 1, Address, City and Postal Code, Country}
% \address[2]{Affiliation 2, Address, City and Postal Code, Country}

% \received{xx xx 202x}
% \finalform{xx xx 202x}
% \accepted{xx xx 202x}
% \availableonline{xx xx 202x}
% \communicated{xx xx}

\author[1]{Gongning {Luo}}
\cortext[cor1]{Corresponding author:}
% \author[1]{Kuanquan {Wang}\corref{cor1}}
% \ead{wangkq@hit.edu.cn}
% \author[1]{Jun {Liu}\corref{cor1}}
% \ead{liujun665@hotmail.com}
\author[1]{Kuanquan {Wang}}
\author[1]{Jun {Liu$^{\ast}$}}
\author[2]{Shuo {Li}}
\author[1]{Xinjie {Liang}}
\author[1]{Xiangyu {Li}}
\author[3]{Shaowei {Gan}}
% \author[1]{Wei {Wang}\corref{cor1}}
% \ead{wangwei2019@hit.edu.cn}
\author[1]{Wei {Wang$^{\ast}$}}
\author[3]{Suyu {Dong}}
\author[1]{Wenyi {Wang}}
\author[4]{Pengxin {Yu}}
\author[4]{Enyou {Liu}}
\author[5]{Hongrong {Wei}}
\author[5]{Na {Wang}}
\author[6]{Jia {Guo}}
\author[6]{Huiqi {Li}}
\author[7]{Zhao {Zhang}}
\author[7]{Ziwei {Zhao}}
% \author[8]{Given-name1 {Surname1}}
% \author[8]{Given-name2 {Surname2}}
\author[8]{Na {Gao}}
\author[9]{Nan {An}}
\author[10]{Ashkan {Pakzad}}
\author[10]{Bojidar {Rangelov}}
\author[11]{Jiaqi {Dou}}  % T8
\author[12]{Song {Tian}}
\author[13]{Zeyu {Liu}} % T9
\author[13]{Yi {Wang}}
\author[14]{Ampatishan {Sivalingam}}
\author[14]{Kumaradevan {Punithakumar}}
% \author[3]{Zhaowen {Qiu}\corref{cor1}}
% \ead{qiuzw@nefu.edu.cn}
\author[3]{Zhaowen {Qiu$^{\ast}$}}
\author[15]{Xin {Gao$^{\ast}$}}

\address[1]{School of Computer Science and Technology, Harbin Institute of Technology, Harbin 150001, China.}
\address[2]{Department of Computer and Data Science and Department of Biomedical Engineering, Case Western Reserve University, Cleveland 44106, USA.}
\address[3]{Institute of Information and Computer Engineering, NorthEast Forestry University, Harbin 150040, China}
\address[4]{Infervision Medical Technology Co., Ltd., Beijing, China}
\address[5]{Sensetime, Shanghai, China}
\address[6]{Beijing Institute of Technology, Beijing, China}
\address[7]{Peking University, Beijing, China}
% \address[8]{Independent Researcher}
\address[8]{Northeastern University, Shenyang, China}
\address[9]{Shanghai Jiao Tong University , Shanghai, China}
\address[10]{University College London (UCL), London, UK}
\address[11]{Center for Biomedical Imaging Research, Tsinghua University, Beijing, China;}
\address[12]{Philips Healthcare, China}
\address[13]{Chongqing University, Chongqing, China}
\address[14]{University of Alberta, Edmonton, Canada}
\address[15]{Computer, Electrical and Mathematical Sciences $\&$ Engineering Division, King Abdullah University of Science and Technology, 4700 KAUST, Thuwal 23955, Saudi Arabia}

\received{xx xx 202x}
\finalform{xx xx 202x}
\accepted{xx xx 202x}
\availableonline{xx xx 202x}
% \communicated{K. Wang, J. Liu, W. Wang, Z. Qiu}

\begin{abstract}
%%%
Efficient automatic segmentation of multi-level (i.e. main and branch) pulmonary arteries (PA) in CTPA images plays a significant role in clinical applications. However, most existing methods concentrate only on main PA or branch PA segmentation separately and ignore segmentation efficiency. Besides, there is no public large-scale dataset focused on PA segmentation, which makes it highly challenging to compare the different methods. To benchmark multi-level PA segmentation algorithms, we organized the first \textbf{P}ulmonary \textbf{AR}tery \textbf{SE}gmentation (PARSE) challenge. On the one hand, we focus on both the main PA and the branch PA segmentation. On the other hand, for better clinical application, we assign the same score weight to segmentation efficiency (mainly running time and GPU memory consumption during inference) while ensuring PA segmentation accuracy. We present a summary of the top algorithms and offer some suggestions for efficient and accurate multi-level PA automatic segmentation. We provide the PARSE challenge as open-access for the community to benchmark future algorithm developments at \url{https://parse2022.grand-challenge.org/Parse2022/}.
%%%%
\end{abstract}

\begin{keyword}
% \MSC 41A05\sep 41A10\sep 65D05\sep 65D17
%% Keywords
\KWD Segmentation\sep Pulmonary artery\sep Multi-level\sep Efficiency
\end{keyword}

\end{frontmatter}

%\linenumbers
 
\section{Introduction}
\label{introduction}
As multi-level vascular systems, pulmonary arteries (PA) are composed of pulmonary trunks, left main pulmonary arteries, right main pulmonary arteries (denoted as main PA), and a large number of lobe segmental branches in the lungs (denoted as branch PA), which transport deoxygenated blood from the heart. PA are associated with many diseases, such as Pulmonary Hypertension (PH)\citep{galie20162015, shahin2022quantitative} and Pulmonary Embolism\citep{konstantinides20202019}. Computed tomography pulmonary angiography (CTPA) is a widely used medical imaging technology for diagnosis, visualization and treatment planning of various PA diseases\citep{galie20162015, konstantinides20202019}. One prerequisite step is to identify and segment multi-level PA structures from CTPA with high accuracy and efficiency\citep{estepar2013vessel12, shahin2022quantitative}. 
% However, manual delineation of the PA in 3D CTPA scans is time-consuming, tedious, poorly reproducible and segmentation results is usually operator-dependent.

% 对分支血管进行分割的必要性
In recent years, more and more studies have shown that the extraction of pulmonary branch blood vessels is of critical significance for the diagnosis and treatment of lung diseases. Shahin et al.\citep{shahin2022quantitative} focused on the size and volume of peel pulmonary vessels and small pulmonary vessels and discovered they are associated with the subtypes of PH. Poletti et al.\citep{poletti2022automated} divided the lung region into three isovolumetric parts, including peripheral, middle and central zones, corresponding to different blood vessel sizes respectively. They analyzed the volume of blood vessels in three areas of normal and diseased lungs (Influenza and Severe Acute Respiratory Syndrome Coronavirus 2), and found significant volume changes in the blood vessels, including branched vessels in the middle and central areas and main vessels in the central area.

\begin{figure*}
	\centering
	\includegraphics[]{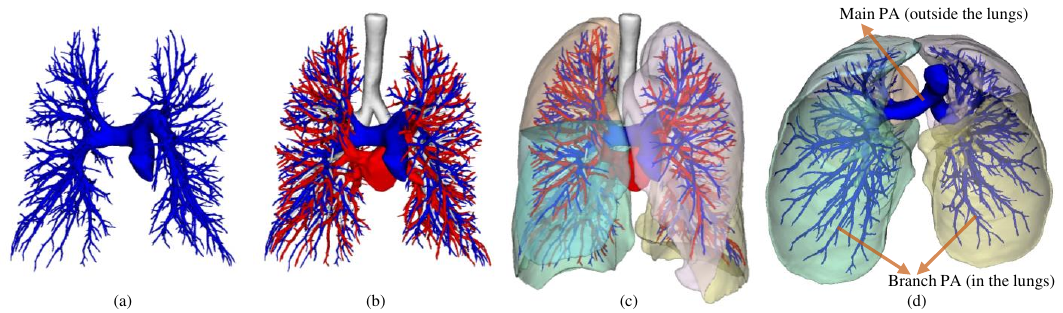}
	\caption[]{Complex pulmonary artery topology structures: (a) the pulmonary arteries (blue); (b) the pulmonary arteries and pulmonary veins (red) are intertwined, while the pulmonary airways (grey) increasing the complex; (c) the pulmonary arteries, pulmonary veins and pulmonary airways with lungs (shade); (d) the pulmonary arteries are divided into two levels: main PA (outside the lungs) and branch PA (inside the lungs).}
	\label{Data_description}
\end{figure*}

However, accurate and efficient PA segmentation remains challenging as shown in Fig. \ref{Data_description}.
(1) \textbf{Complex topology structures:} From the main PA to the branch PA, there is a large number of bifurcations (Fig. \ref{Data_description} (b)), especially in the distal part of the branch. The bifurcation process along the artery is accompanied by the reduction of the cross-sectional area of blood vessels and the increase of vascular density, indicating the increasing difficulty of PA segmentation. The PA also have similar topological structures to the lung's airways and pulmonary veins, particularly the pulmonary veins, which are often intertwined with the pulmonary arteries as shown in Fig. \ref{Data_description} (a).
(2) \textbf{Large and fine regions of interest (ROI):} The PA region that needs to be segmented is large, and both the main PA and the fine branch PA need to be accurately segmented. Thus, it is difficult to ensure segmentation accuracy and efficiency simultaneously.
(3) \textbf{Inter-class imbalance:} There is a serious imbalance between the foreground (PA region) and the background, especially at the end of branch vessels.
(4) \textbf{Intra-class imbalance:} The main PA have more voxels in the CT image and is relatively easy to segment. Nevertheless, the branch PA have few voxels and tend to break or miss during segmentation tasks.
(5) \textbf{Large individual variability:} Different individuals naturally have different appearances and geometric shapes of PA, coupled with disease factors, which further lead to variability. The use of contrast agents and the setting of imaging instrument parameters may introduce artifacts and inherent image noise and will increase imaging variability.
(6) \textbf{Difficulty in annotation:} It is difficult to completely and accurately label all pulmonary arteries because of the extremely complex topology. The annotations may be noisy.

During the past decades, to the best of our knowledge, few public datasets have been proposed to focus on the segmentation of PA. VESSEL12\citep{estepar2013vessel12} challenge targeted for automated lung vessel segmentation in computed tomography (CT) scans and published only three annotated samples. CARVE14\citep{charbonnier2015automatic} targeted to pulmonary artery-vein separation and classification with 55 non-contrast CT scans, which were from the ANODE challenge\citep{van2010comparing}. However, the manual annotations of both VESSEL12 and CARVE14 datasets were performed only on points of interest and not the entire vessel tree, which was hard for training in deep learning. Tan et al.\citep{tan2021automated} reviewed automated pulmonary vascular segmentation algorithms from the International Symposium on Image Computing and Digital Medicine 2020 challenge, including 16 sets of CT plain scan images and 16 sets of computed tomography angiography (CTA) enhanced images. However, no dataset download link is provided. Therefore, a large-scale dataset focused on PA segmentation is missing.

To address these limitations, we organized the \textbf{P}ulmonary \textbf{AR}tery \textbf{SE}gmentation (PARSE) challenge in conjunction with the International Conference on Medical Image Computing and Computer Assisted Intervention (MICCAI) 2022, held in Singapore. Participants were required to develop automatic main and branch PA segmentation algorithms, and both segmentation accuracy and efficiency were evaluated for the final ranking.

In this paper, we introduce an overview of the PARSE challenge and discuss the top algorithms. The main contributions are summarized as follows:
\begin{itemize}
  \item We organize the first PA segmentation challenge that focuses on both main PA and branch PA segmentation to achieve high accuracy and efficiency simultaneously.
  \item We analyze the submitted algorithms and summarize the results of the top teams.
  \item We present various algorithmic techniques for improving the accuracy and efficiency of PA segmentation and provide recommendations. 
\end{itemize}

\begin{table*}[]
  \centering
  \caption{Images information (numbers, gender, age, vendor, center, and physical sizes) of the training, validation, and testing sets. All the slice thickness is 1 mm.}
  \label{tab:data}
  \setlength{\tabcolsep}{1.5mm}
  \begin{tabular}{cccccccccc}
    \hline \toprule
     & Samples & Gender & \multirow{2}{*}{Age} & \multicolumn{3}{c}{SIEMENS} & PHILIPS & \multirow{2}{*}{Slice Nums} & Pixel Spacing \\ \cmidrule(r){5-7} 
             & Nums & (F/M) &             & Center1 & Center2 & Center3 & Center4 &              &  (mm)                     \\ \hline
    Training & 100  & 79/21 & 29-71 (56.3) & 81      & 16      & 3       &         & 228-390 (303) & 0.5039-0.9238 (0.6740) \\
    Validation    & 30   & 25/5  & 29-70 (54.2) & 25      & 3       & 2       &         & 228-397 (295) & 0.5469-0.7891 (0.6651) \\
    Testing  & 73   & 52/21 & 23-73 (54.4) & 61      & 9       &         & 3       & 229-408 (312) & 0.5039-0.8770 (0.6906) \\ \hline \toprule
  \end{tabular}
\end{table*}

\section{Methods}
\label{methods}
\subsection{Challenge dataset}
PARSE challenge provided CTPA image sets from 203 subjects, which had been diagnosed with pulmonary nodular disease. The in-plane size and resolution were $512\times512$ pixels and $0.5\sim0.95$ mm/pixel, respectively. The slices number and thickness were $228\sim408$ and 1 mm, respectively. All the data had been anonymized. The dataset division and detailed parameters are shown in Table \ref{tab:data}. Among them, age, slice numbers and pixel spacing are shown with the minimum, maximum and mean values. The data were obtained from four centers (Harbin Medical University Cancer Hospital, Chinese PLA General Hospital, Hongqi Hospital Affiliated to Mudanjiang Medical University, and the First Affiliated Hospital of Jiamusi University) and two device manufacturers (SIEMENS and PHILIPS).

The annotating work involved the participation of 10 experts with more than 5-year clinical experience, who were divided into two groups. Each group included 5 experts and each scan was annotated by 5 experts. Annotations were performed using MIMICS software semi-automatically based on region growing method. Firstly, window width and window level were adjusted to show the PA structures clearly. Secondly, seed points were selected iteratively and manually to get a coarse mask. Thirdly, all the experts finetuned the mask results of PA structure and checked the annotation mutually. Finally, the voting method was applied to merge the 5 annotations for each scan. 

The data were split into three sets, i.e., the training set, the validation set and the testing set with case number 100, 30 and 73, respectively. Participants could access the training images with corresponding annotations and the validation images without annotations. The validation set annotations and the testing images with corresponding annotations were held by organizers.

\subsection{Challenge organization}
When organizing the PARSE challenge and writing this paper, we followed the BIAS (Biomedical Image Analysis ChallengeS) guideline\citep{maier2020bias}. The PARSE challenge included a training phase, a validation phase and a testing phase. During the training phase, participants could develop and cross-validate a robust segmentation algorithm with training images and corresponding masks after their applications were approved. During the validation phase, only three segmentation results of the validation images could be uploaded to the official website to validate the algorithms on each day and the scores could be shown automatically. Once the participants developed their methods, they submitted their Docker container that could produce segmentation results with a specific command before the end of the testing phase. Only one working Docker container was confirmed to produce the final testing results for each team. A short paper that included sufficient method details was submitted by each ranking team.

For a fair comparison, both pre-trained models and additional training data were not allowed in this challenge. All ranking teams were awarded a specific certificate and the top 3 teams were awarded cash. Their final results were exhibited publicly. The top 10 teams were invited to show their excellent algorithms at the MICCAI challenges conference and to be co-authors of the challenge review paper.

\subsection{Evaluation metrics and ranking scheme}
Since PA targets have a large ROI and a fine structure to segment, it is difficult to ensure segmentation accuracy while ensuring segmentation efficiency. However, both accuracy and efficiency metrics are needed for clinical application\citep{ma2022fast}. Therefore, motivated by FLARE challenge\citep{ma2022fast}, we adopted accuracy-oriented and efficiency-oriented evaluation metrics. Specifically, for accuracy-oriented metrics, the widely used Dice Similarity Coefficient (DSC) and 95\% Hausdorff Distance (HD95) were selected. For efficiency-oriented metrics, running time (RT) and maximum used GPU memory (GPU) were selected, similar to the FLARE challenge.

As shown in Fig. \ref{Data_description}, the segmentation target of PARSE could be roughly divided into two levels, i.e., main PA and branch PA. It's unsuitable to treat the two of them equally, because the branch PA is much more difficult to extract. Therefore, we introduced multi-level accuracy metrics for multi-level PA. Specifically, we adopted two-level DSC and HD95 metrics: (1) level one: DSC and HD95 metrics for branch PA, which is inside the lung area; (2) level two: DSC and HD95 metrics for main PA, which is outside the lung area. It is worth noting that we give more weight to the branch PA segmentation metrics (level one) because the thinner structures are more difficult for segmentation and more critical for clinical application\citep{shahin2022quantitative}. To be specific, the weights of level one and level two are 80\% and 20\% for the final scores, respectively.

After we got all the teams' final scores, we implemented the following ranking scheme:
\begin{itemize}
  \item Step 1. The weighted DSC and HD95 scores of all testing cases of each team were calculated separately and then averaged to obtain the mean DSC and HD95 scores of each team. And then, the average RT and GPU were computed for each team.
  \item Step 2. The weighted DSC, weighted HD95, RT and GPU average scores were ranked separately for all teams.
  \item Step 3. These four rankings were averaged for each team.
  \item Step 4. Based on the average rankings in step 3, the final ranking was determined. Tied them if the average rankings of two teams were equal.  
\end{itemize}

\section{Results}
\subsection{Challenge participants and submissions}
\begin{figure}
	\centering
	\includegraphics[]{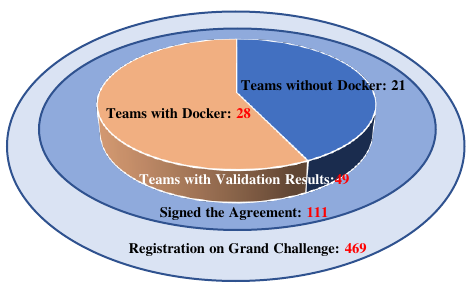}
	\caption[]{Summary of PARSE challenge participants and submissions. There were 469 teams registering on the official grand-challenge webpage and 111 of them were approved before the end of the training phase. Finally, 49 teams submitted validation results and 28 teams submitted Docker containers with 25 qualified results.}
	\label{Participants}
\end{figure}

The official PARSE challenge was located on the grand-challenge platform\footnote{\url{https://parse2022.grand-challenge.org/Parse2022/}}. Participants should sign the challenge rule agreement and send it to the official mailbox. Fig. \ref{Participants} shows information about participants and submissions. Specifically, we received more than 400 applications from over 30 countries on the grand-challenge webpage and 111 teams were approved with a complete signed application form. During the validation phase, 49 teams submitted validation results. During the testing phase, 28 teams submitted Docker containers. But 3 Docker containers couldn't work properly. Finally, we got 25 qualified submissions.

\begin{table*}
  \centering
  \caption{Summary of the benchmark methods of top ten teams.}
  \label{tab:top10}
  \setlength{\tabcolsep}{1.5mm} % 此处将表格按照页面宽度调整
  
  {\begin{tabular}[]{llll}
    \hline
    \toprule
    Team                 & Framework                  & Network                                     & Highlighting methods                                                          \\ \hline
    \multirow{5}{*}{IMT} & \multirow{5}{*}{One-stage} & \multirow{5}{*}{U-Net with residual blocks} & Lung area extracted by threshold method; \\ 
            & & & Modified General Union Loss and weighted\\
            & & & cross-entropy loss;\\ 
            & & & Hard example mining strategy;\\ 
            & & & Parameter averaging.\\ \hline
    \multirow{4}{*}{ST}  & \multirow{4}{*}{One-stage} & \multirow{4}{*}{nnU-Net}                    & Lung area extracted by threshold method; \\
            & & & Suitable data augmentation; \\
            & & & Over-sampling mechanism; \\ 
            & & & Selected test-time view augmentation. \\ \hline
    \multirow{3}{*}{BIT} & \multirow{3}{*}{Two-stage} & \multirow{3}{*}{U-Net with residual blocks} & Lung area extracted by trained network; \\
            & & & Label noise tackling; \\
            & & & Weighted patch assembling. \\ \hline
    \multirow{3}{*}{PKU} & \multirow{3}{*}{One-stage} & \multirow{3}{*}{Skeleton-aware U-Net}       & Lung area extracted by threshold method and\\
            & & & connected components discovering; \\ 
            & & & Appending a skeleton decoder. \\ \hline
    \multirow{3}{*}{IR}  & \multirow{3}{*}{One-stage} & \multirow{3}{*}{U-Net}                      & Adaptive sliding window strategy; \\  
            & & & Removing air operation; \\ 
            & & & Deep supervision.  \\ \hline
    \multirow{3}{*}{NEU} & \multirow{3}{*}{One-stage} & \multirow{3}{*}{Dual U-Net}                 & Lung area extracted by threshold method; \\  
            & & & Two U-Net structures with different input sizes and \\
            & & & down-sampling times.      \\ \hline
    \multirow{2}{*}{UCL} & \multirow{2}{*}{One-stage} & \multirow{2}{*}{nnU-Net}                    & Morphological closure; \\  
            & & & Models ensemble. \\ \hline
    \multirow{2}{*}{THU} & \multirow{2}{*}{One-stage} & \multirow{2}{*}{U-Net}                      & Two training stages;                                                          \\  
            & & & Fine-tune with clDice. \\ \hline
    CQU                  & Two-stage                  & U-Net                                       & Coarse-to-fine strategy.                                                      \\ \hline
    IITBHU               & One-stage                  & nnU-Net                                     & Image-orientation strategy.                                                   \\ \hline \toprule
  \end{tabular}}
\end{table*}

\subsection{Algorithms summarization of top ten teams}
We highlight the key points of the top ten algorithms in Table \ref{tab:top10}. Their key strategies to enhance segmentation accuracy and efficiency were demonstrated as follows.

\subsubsection{T1: Infervision Medical Technology Co., Ltd (IMT)}
\begin{figure}
	\centering
	\includegraphics[width=8.5cm]{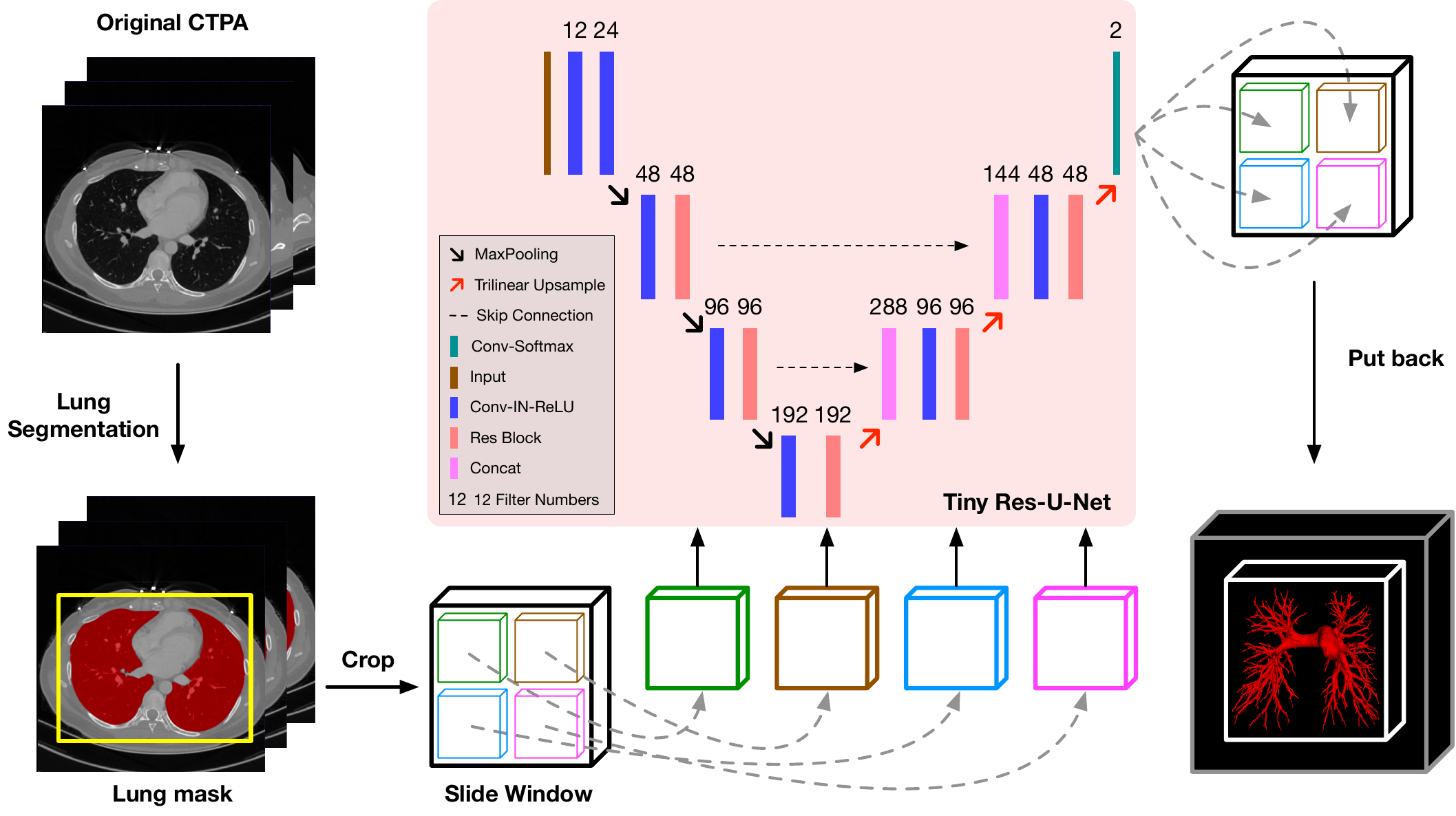}
	\caption[]{Framework pipeline for the Top 1 team. A lightweight model called Tiny Res-U-Net was designed.}
	\label{Top1}
\end{figure}

IMT developed an efficient U-Net\citep{unet2015} framework with residual blocks\citep{residual2016} as shown in Fig \ref{Top1}. In the training phase, IMT designed three aspects (i.e. data, model and loss function) to achieve the most optimal trade-off between segmentation accuracy and efficiency. For data, IMT proposed the following special design: (1) use two windows to clip the intensity of the images and construct them as a two-channel input; (2) use the hard example mining strategy\footnote{After prediction probability map was achieved from a trained model, positive class voxels with predicted probabilities between 0.3 and 0.5, and negative class voxels with predicted probabilities between 0.5 and 0.8 were defined as hard samples} to sample the training data so that the difficult-to-distinguish regions can be more concerned; (3) apply lung region segmentation by threshold method to extract ROI, which will help both inference speed and accuracy. For the model, IMT built a lightweight model called Tiny Res-U-Net as shown in Fig \ref{Top1} to perform PA segmentation. Specifically, only three downsamplings were performed and no skip connection was used for the original resolution in Tiny Res-U-Net. Moreover, the number of channels is small, so both RT and GPU memory consumption can be guaranteed. Furthermore, IMT experimentally found that deeper or wider networks didn't improve performance. For the loss, IMT modified the General Union Loss\citep{zheng2021alleviating} and used a weighted cross-entropy (CE) loss as an auxiliary loss, which only calculates the loss of voxels in the lung. During inference, IMT converted the model and data to half-precision, which could effectively reduce GPU memory consumption with almost no impact on accuracy. In addition, IMT used parameter averaging for the model ensemble. IMT found that although it performed slightly worse than ensembled predictions of multiple models, it wouldn't introduce additional computation. Further, a connected region analysis was performed to reduce false positive results.

\subsubsection{T2: SenseTime (ST)}
ST modified nnU-Net\citep{nnunet} with multiple aspects. (1) Data augmentation: improve the data augmentation of the original nnU-Net with elastic deformation and global intensity perturbations to better generalize the PARSE data. To be specific, the probability of elastic deformation was setted as 0.2 with $\alpha \in$ [0, 1000] and $\sigma \in$ [9, 13], while the probability of global intensity perturbations following Gaussian distribution N(0, 0.1) was setted as 0.15. (2) Over-sampling mechanism: sampling probability for branch PA was improved to 0.7 to make the network give more attention to the branch. (3) Specific test-time augmentation (TTA) view: the segmentation performance of each augmented view (i.e., flipping the explicit axis) was evaluated and only four key TTA views were retained to balance accuracy and efficiency. In addition, lung area was extracted by threshold method to speed up the inference and reduce false positives.

\subsubsection{T3: Beijing Institute of Technology (BIT)}
BIT proposed a 3D Res-Unet that used U-Net backbone with residual blocks\citep{residual2016}. BIT adopted instance normalization\citep{loshchilov2018decoupled} (IN) instead of batch normalization (BN) as the former was more friendly to small batch size. To alleviate the label noise in the main PA, the predicted results of five-fold cross-validation and the original ground truth were assembled to recompose the label. The recomposed label was then used to train the final network. A neural network trained with lung masks extracted by the traditional method was used to crop the lung area during inference. BIT believed it was faster than traditional method for extracting the lung region. Because reading and writing Nifty file (.nii.gz) is time-consuming, BIT used asynchronous threads to read and write along with the main processing to improve inference efficiency.

\begin{figure*}
  \centering
  \includegraphics[width=14cm]{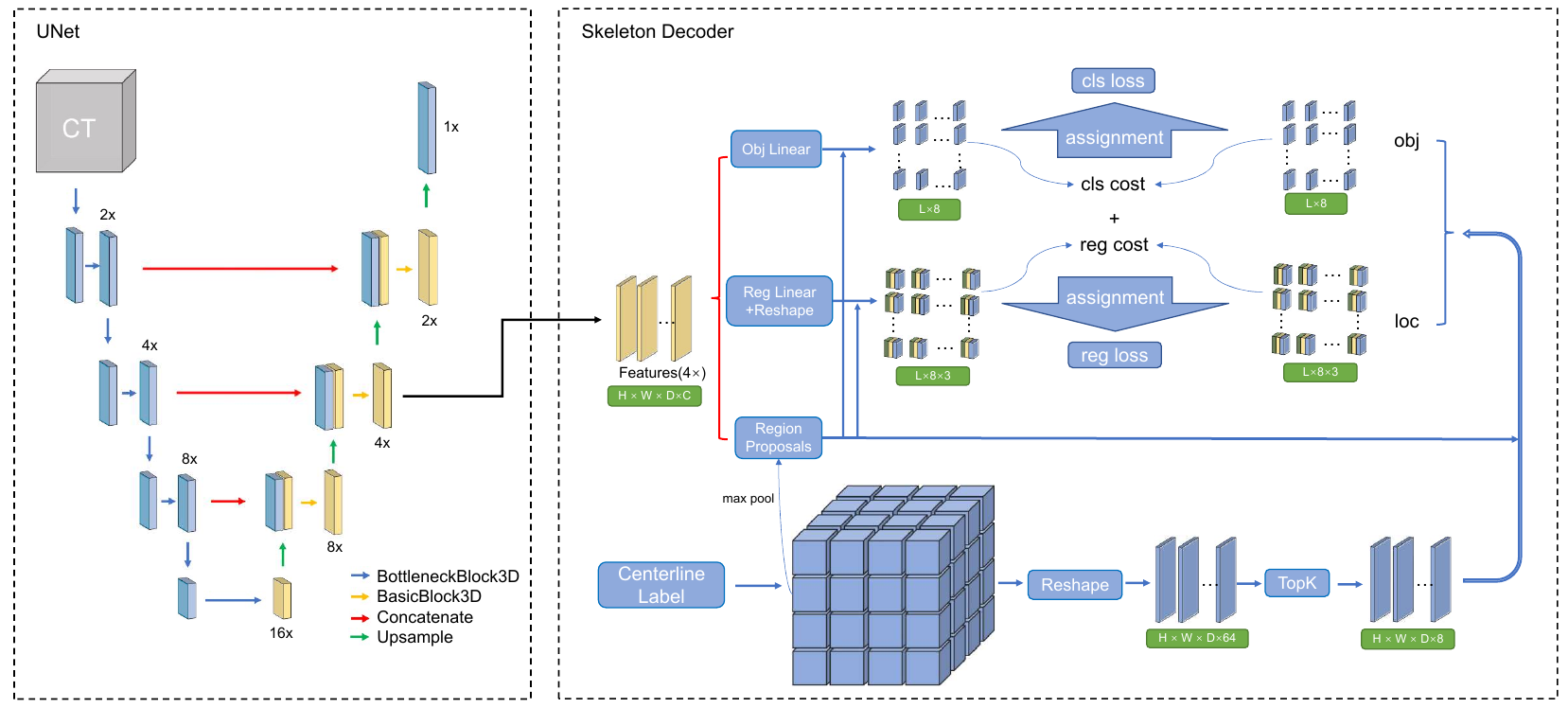}
  \caption[]{Framework overview of the T4 team. The model consists of a U-Net architecture to perform segmentation and a skeleton decoder used in the training stage to predict vessel skeletons.}
  \label{Top4network}
\end{figure*}

\subsubsection{T4: Peking University (PKU)}
Compared with the PA masks, PKU deemed that the skeletons of PA had optimal advantages: (1) skeleton structures were not affected by the thickness of vessels; (2) skeletons could reflect the topology structures of PA more accurately. Therefore, PKU appended a skeleton decoder to the Res-Unet\citep{zhang2018road} (Fig. \ref{Top4network}) for extracting vessel skeletons in the training phase and discarded it in the inference stage. Specifically, the input of the skeleton decoder was 4 $\times$ downsampled feature maps from the Res-Unet decoder. Three linear layers were introduced to predict the presence of PA at each position, objectness scores and coordination offsets relative to the region center points with Focal Loss\citep{lin2017focal}, Binary CE loss and L1 loss, respectively. Besides, the lung mask was extracted through the threshold method and connected components analysis to eliminate false positive predictions effectively and efficiently.

\subsubsection{T5: Independent Researcher (IR)}
IR adopted a U-Net\citep{unet2015} as the backbone and referred to the implemented experience of nnU-Net\citep{nnunet}. An adaptive sliding window strategy and removing air from images were used before feeding them to the network to reduce the number of patches and increase efficiency. Moreover, deep supervision was implemented to alleviate the false positive problem.
% removing air 是什么意思？具体怎么做的？adaptive sliding window strategy是怎么做的？

\subsubsection{T6: Northeastern University (NEU)}
NEU developed a 3D dual U-Net that was composed of two 3D U-Net\citep{cciccek20163d} structures with different input sizes ([128, 160, 192] and [112, 128, 160], respectively) and down-sampling times (5 and 4, respectively). NEU believed U-Net with a large input size was more effective at extracting global features, thereby reducing false positives, while U-Net with a small input size had an advantage at extracting local features, thereby being useful for segmenting the foreground. To be specific, the lung region was first extracted through the threshold method and ROI was achieved by cropping images based on the lung box. Then, two sizes of patches ([128, 160, 192] and [112, 128, 160]) were fed to the two networks, respectively. At last, the outputs of the last layer of both networks were element-wise added to predict the PA mask. Besides, the lung mask was also used for filtering false positives during inference to improve accuracy and efficiency.
% 两个网络输入尺寸和下采样次数分别是多少？（图中看起来下采样次数一样）

\subsubsection{T7: University College London (UCL)}
UCL leveraged the nnU-Net\citep{nnunet} to create multiple models and ensembled their predictions to produce final outputs. Specifically, there existed four models, i.e., 2D only, 3D low resolution, 3D full resolution and 3D cascaded models. In addition, morphological closure of the final output with intensity thresholding was adopted to improve connectivity.
% 组装了3个模型，还是4个模型？

\subsubsection{T8: Tsinghua University (THU)}
THU developed a 3D U-Net\citep{unet2015} with two training stages. The model was first trained with Dice loss and CE loss to quickly locate the PA area. And then, the model was fine-tuned with the clDice loss\citep{shit2021cldice} to focus on the center-line of the PA ground truth.

\subsubsection{T9: Chongqing University (CQU)}
CQU proposed a coarse-to-fine framework based on U-Net\citep{unet2015}. The coarse and fine models shared the same network architecture with different input sizes, i.e., 160 $\times$ 160 $\times$ 160 and 96 $\times$ 96 $\times$ 96, respectively. The images were first resized and then fed to the coarse model to locate the ROI. Subsequently, the images were cropped according to the outputs of the coarse model and fed to the fine model to achieve fine segmentation.

\subsubsection{T10: University of Alberta - Indian Institute of Technology (BHU) Varanasi (UA-IITBHU)}
UA-IITBHU developed an image-orientation strategy based on nnU-Net\citep{nnunet}. During the training phase, the images and corresponding labels were first oriented to be parallel to the coordinate axes by removing the position and orientation information from the header file of NIFTI images, and then they were sent to train the network according to the standard setting of the nnU-Net framework. During inference, the images were first oriented as described above and then sent to the trained model to get the prediction, followed by a reorientation to match the original images to get the final predictions.

\subsection{Evaluation results and ranking analysis}
Table \ref{tab:all_results} shows the average DSC and HD95 for main, branch and weighted PA, and the average RT and GPU metrics of the 25 qualified teams. In the next subsections, we present the results analysis of the DSC and HD95 metrics in Table \ref{tab:all_results} by dot- and boxplots visualization and statistical significance maps (Fig. \ref{DSC_blox} and Fig. \ref{HD95_blox}), based on the ChallengeR\citep{wiesenfarth2021methods} that is a professional tool for challenge results analysis and visualization. Statistical significance maps are analyzed using the one-sided Wilcoxon signed rank test at a significance level of 5\%, which is used in many challenge results analysis\citep{wiesenfarth2021methods,ma2022fast}. Then, we perform comparative analysis for RT and GPU metrics. For a concise and unambiguous description, we mainly focus on the top 20 teams.

\begin{table*}[]
  \centering
  \caption{Quantitative evaluation results of the qualified 25 teams in terms of (mean ± standard deviation) Dice Similarity Coefficient (DSC), 95\% Hausdorff Distance (HD95), Running time per case (RT), MAX GPU memory consumption (GPU).}
  \label{tab:all_results}
  \begin{tabular}{ccccccccc}
  \hline
  \toprule
  \multicolumn{1}{c}{\multirow{2}{*}{Teams}} &
    \multicolumn{3}{c}{DSC (\%) $\uparrow$} &
    \multicolumn{3}{c}{HD95 (mm) $\downarrow$} &
    \multicolumn{1}{c}{RT $\downarrow$} &
    \multicolumn{1}{c}{GPU $\downarrow$} \\    
    \cmidrule(r){2-4}
    \cmidrule(r){5-7}
  \multicolumn{1}{c}{} &
    \multicolumn{1}{c}{Main PA} &
    \multicolumn{1}{c}{Branch PA} &
    \multicolumn{1}{c}{Weighted PA} &
    \multicolumn{1}{c}{Main PA} &
    \multicolumn{1}{c}{Branch PA} &
    \multicolumn{1}{c}{Weighted PA} &
    \multicolumn{1}{c}{(s/case)} &
    \multicolumn{1}{c}{(MB)} \\ 
    \hline
    T1 & $89.70\pm6.46$ & $\textbf{77.19}\pm\textbf{8.32}$ & $\textbf{79.69}\pm\textbf{6.95}$ & $7.08\pm6.62$ & $\textbf{4.80}\pm\textbf{4.12}$ & $5.26\pm3.54$ & 7.92  & \textbf{1674} \\
    T2 & $\textbf{91.71}\pm\textbf{4.00}$ & $76.67\pm8.39$ & $\textbf{79.68}\pm\textbf{6.94}$ & $\textbf{4.83}\pm\textbf{4.97}$ & $\textbf{4.74}\pm\textbf{4.17}$ & $\textbf{4.75}\pm\textbf{3.66}$ & 18.61  & 3658 \\
    T3 & $\textbf{91.57}\pm\textbf{4.25}$ & $\textbf{76.86}\pm\textbf{7.94}$ & $\textbf{79.80}\pm\textbf{6.61}$ & $\textbf{4.99}\pm\textbf{5.11}$ & $5.46\pm4.77$ & $5.36\pm4.11$ & 14.53  & 6300 \\
    T4 & $89.50\pm6.58$ & $75.53\pm8.85$ & $78.33\pm7.49$ & $6.76\pm6.42$ & $4.92\pm4.00$ & $5.29\pm3.57$ & \textbf{6.63}  & 3326 \\
    T5 & $89.98\pm4.94$ & $\textbf{76.76}\pm\textbf{7.88}$ & $79.40\pm6.38$ & $7.74\pm7.72$ & $5.70\pm4.87$ & $6.10\pm3.72$ & 63.11  & 2828 \\
    T6 & $90.05\pm5.65$ & $76.76\pm8.15$ & $79.42\pm6.89$ & $8.38\pm11.32$ & $5.44\pm4.61$ & $6.03\pm4.53$ & 25.55  & 8176 \\
    T7 & $90.49\pm5.13$ & $76.57\pm8.29$ & $79.36\pm6.80$ & $6.37\pm6.30$ & $\textbf{4.56}\pm\textbf{3.85}$ & $\textbf{4.92}\pm\textbf{3.35}$ & 263.24  & 3658 \\
    T8 & $\textbf{90.83}\pm\textbf{4.68}$ & $76.25\pm8.29$ & $79.17\pm6.80$ & $9.19\pm19.91$ & $5.17\pm4.13$ & $5.98\pm5.22$ & 57.57  & 4120 \\
    T9 & $90.29\pm5.04$ & $76.53\pm8.33$ & $79.28\pm6.87$ & $6.50\pm6.13$ & $5.36\pm4.51$ & $5.58\pm3.86$ & 55.71  & 9303 \\
    T10 & $90.45\pm5.20$ & $76.10\pm8.71$ & $78.97\pm7.16$ & $6.18\pm6.30$ & $4.93\pm4.34$ & $5.18\pm3.60$ & 380.57  & 3472 \\
    T11 & $90.08\pm5.37$ & $75.71\pm8.68$ & $78.58\pm7.29$ & $9.19\pm15.00$ & $4.94\pm3.72$ & $5.79\pm4.19$ & 88.93  & 3658 \\
    T12 & $89.69\pm5.30$ & $75.46\pm8.40$ & $78.31\pm6.94$ & $31.65\pm57.21$ & $5.91\pm4.55$ & $11.06\pm11.81$ & \textbf{7.74}  & 4298 \\
    T13 & $90.12\pm4.82$ & $74.30\pm9.06$ & $77.46\pm7.53$ & $18.43\pm40.91$ & $5.52\pm4.24$ & $8.10\pm9.07$ & 51.14  & 2800 \\
    T14 & $90.55\pm4.63$ & $68.86\pm9.70$ & $73.20\pm7.97$ & $6.24\pm6.17$ & $17.26\pm6.32$ & $15.06\pm5.25$ & 8.11  & \textbf{1894} \\
    T15 & $90.55\pm5.13$ & $76.29\pm8.58$ & $79.15\pm7.12$ & $\textbf{6.05}\pm\textbf{6.58}$ & $4.71\pm4.05$ & $\textbf{4.98}\pm\textbf{3.46}$ & 218.82  & 10242 \\
    T16 & $90.34\pm4.86$ & $74.98\pm8.78$ & $78.05\pm7.17$ & $6.50\pm6.14$ & $5.18\pm4.18$ & $5.44\pm3.58$ & 116.40  & 4988 \\
    T17 & $89.87\pm5.04$ & $70.00\pm9.79$ & $73.98\pm8.08$ & $6.96\pm7.32$ & $8.81\pm7.09$ & $8.44\pm5.77$ & \textbf{4.21}  & 8230 \\
    T18 & $90.19\pm4.43$ & $71.30\pm8.83$ & $75.08\pm7.28$ & $6.13\pm5.56$ & $7.84\pm3.09$ & $7.50\pm2.74$ & 224.22  & 2328 \\
    T19 & $89.74\pm4.85$ & $70.41\pm7.91$ & $74.28\pm6.63$ & $7.20\pm6.04$ & $15.87\pm4.34$ & $14.13\pm3.69$ & 172.93  & 2640 \\
    T20 & $89.84\pm4.85$ & $75.29\pm7.52$ & $78.20\pm6.19$ & $20.65\pm37.27$ & $8.28\pm4.74$ & $10.75\pm7.58$ & 80.11  & 9424 \\
    T21 & $89.80\pm5.24$ & $74.76\pm8.56$ & $77.77\pm7.14$ & $11.77\pm27.81$ & $6.10\pm4.24$ & $7.24\pm6.32$ & 76.69  & 12288 \\
    T22 & $90.55\pm4.94$ & $76.05\pm8.31$ & $78.95\pm6.89$ & $9.14\pm18.45$ & $5.22\pm4.26$ & $6.00\pm5.06$ & 3178.75  & \textbf{0} \\
    T23 & $80.65\pm9.56$ & $62.53\pm7.19$ & $66.15\pm6.42$ & $140.75\pm69.78$ & $11.24\pm2.84$ & $37.15\pm14.5$ & 48.22  & 21996 \\
    T24 & $82.65\pm25.41$ & $69.37\pm22.35$ & $72.02\pm22.61$ & $-$ & $-$ & $-$ & 419.20  & 3349 \\
    T25 & $85.79\pm18.82$ & $69.78\pm16.4$ & $72.98\pm16.52$ & $65.12\pm69.46$ & $-$ & $-$ & 111.50  & 9784 \\ 
    \hline \toprule
  \end{tabular}
\end{table*}

\subsubsection{DSC metric analysis}
\begin{figure*}
  \centering
  \includegraphics[width=16cm]{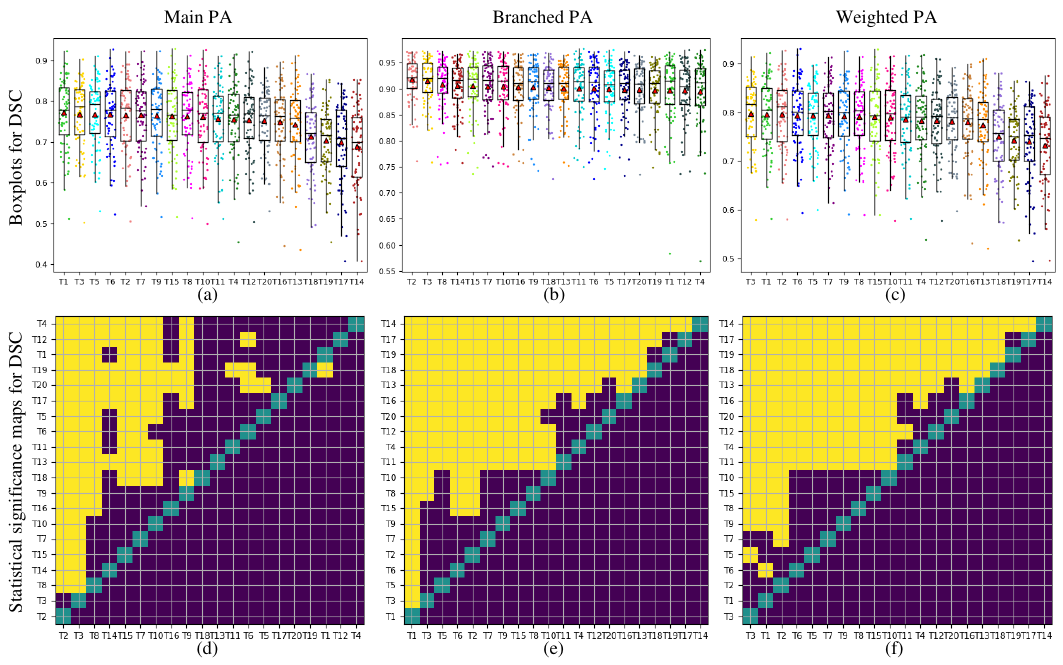}
  \caption[]{Dot- and boxplot visualization (the first row) and statistical significance maps (the second row) for the DSC metric of the top 20 teams. (a) and (d), (b) and (e), and (c) and (f) are the results for main PA, branch PA and weighted PA, respectively. In each sub-figure, the teams are arranged from left to right in accordance with the corresponding mean score. In the statistical significance maps, light red shading indicates that the DSC scores of the teams on the x-axis are significantly superior to the scores of the teams on the y-axis based on one-sided Wilcoxon signed rank test (\emph{p}-value \textless 5\%) and dark red shading indicates they are not significantly superior.}
  \label{DSC_blox}
\end{figure*}

As shown in Table \ref{tab:all_results} and Fig. \ref{DSC_blox} (a - c), all the teams obtained better DSC scores for main PA than those for branch PA. After being weighted, the balanced scores biased towards the branch PA were obtained, as a result of the higher weight that was given to the branch PA. Similarly, from the Table \ref{tab:all_results}, T2, T3 and T8 have obtained top 3 DSC performance for main PA, while T1, T3 and T5 have achieved top 3 DSC performance for branch PA. Therefore, the teams that achieved better DSC performance for main PA probably wouldn't score as well in branch PA, and vice versa. It's demonstrated that it is necessary to evaluate multi-level PA (main and branch PA) separately. 

The statistical significance analysis (Fig. \ref{DSC_blox} (d)) showed that the DSC score of T2 for main PA had no significant difference compared to T3. However, T2 and T3 both obtained DSC scores for main PA that were significantly superior to the scores of the other teams. For branch PA, the statistical significance analysis (Fig. \ref{DSC_blox} (e)) shows that the highest score to belong to T1 was statistically significantly higher than those of all other teams, while the best scores to belong to T3 and T5 were significantly higher than those of other teams in part. After being weighted, the DSC scores of the top 3 (T3, T1 and T2) did not differ notably from each other, while they were superior to most of the scores of other teams with statistical significance (Fig. \ref{DSC_blox} (f)). Therefore, it is meaningful to comprehensively evaluate the multi-level PA (main and branch PA) in a weighted method.

Specifically, team T1 obtained the best DSC score (Table \ref{tab:all_results}) for branch PA and was significantly superior to all the other teams (Fig. \ref{DSC_blox} (e)). The error map of team T1 in Fig. \ref{visualization} demonstrates that there is the least segmentation error in the branch PA region for T1. However, T1 did not perform well in scoring on the main PA(Table \ref{tab:all_results}), especially in the PA root area (referring to the error map in Fig. \ref{visualization}). In addition, T2 earned the best DSC score (Table \ref{tab:all_results}) for the main PA and was significantly outperforming most teams (Fig. \ref{DSC_blox} (d)). The error map of T2 in Fig. \ref{visualization} shows that T2 has the least segmentation error in the main PA region.

\subsubsection{HD95 metric analysis}
\begin{figure*}
  \centering
  \includegraphics[width=16cm]{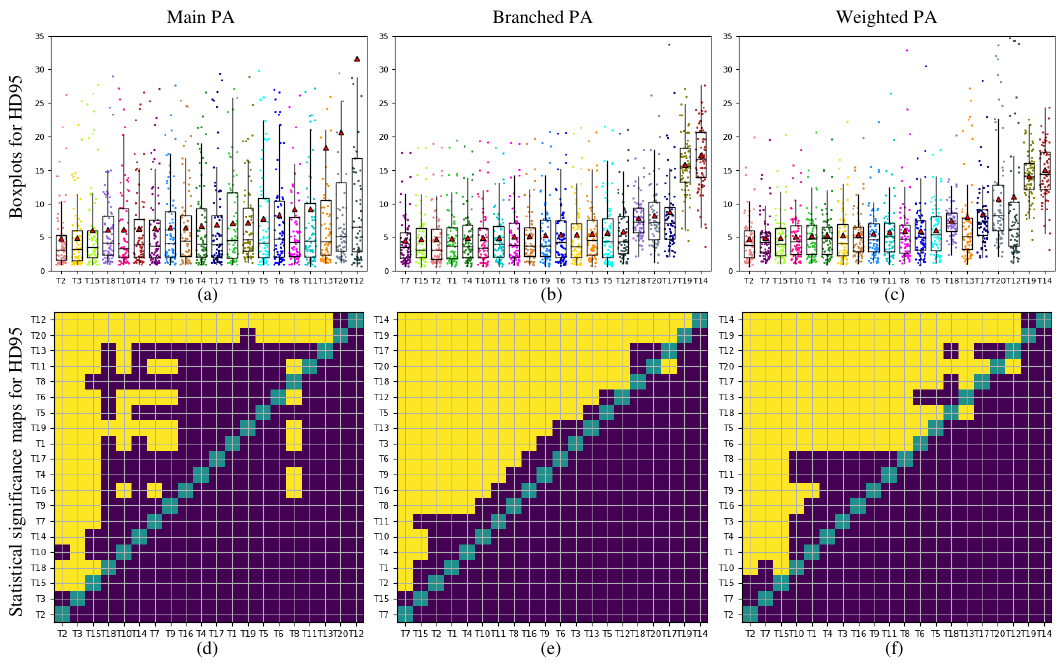}
  \caption[]{Dot- and boxplot visualization (the first row) and statistical significance maps (the second row) for the HD95 metric of the top 20 teams. The illustration in this figure is similar to that in the last figure, except that this figure uses the HD95 metric.}
  \label{HD95_blox}
\end{figure*}

As shown in Table \ref{tab:all_results} and Fig. \ref{HD95_blox} (a - c), unlike the DSC scores, most of the teams obtained better HD95 scores for branch PA, which showed lower average values with lower dispersion, than those for main PA. This result could be attributed to the fact that the structure of branch PA was full of lungs and the calculation of HD95 was confined to the interior of the lungs. Besides, the HD95 scores of the top 3 teams for main, branch and weighted PA were [T2, T3, T10], [T7, T2, T1] and [T2, T7, T10], respectively. Therefore, similar to teams achieving high scores for DSC, each team did not always obtain high scores for HD95 on both main PA and branch PA. Moreover, the top teams for DSC and HD95 scores were remarkably different. Consequently, DSC and HD95 are complementary metrics and it is necessary to introduce both of them to evaluate the algorithms.

\begin{figure*}
  \centering
  \includegraphics[width=17cm]{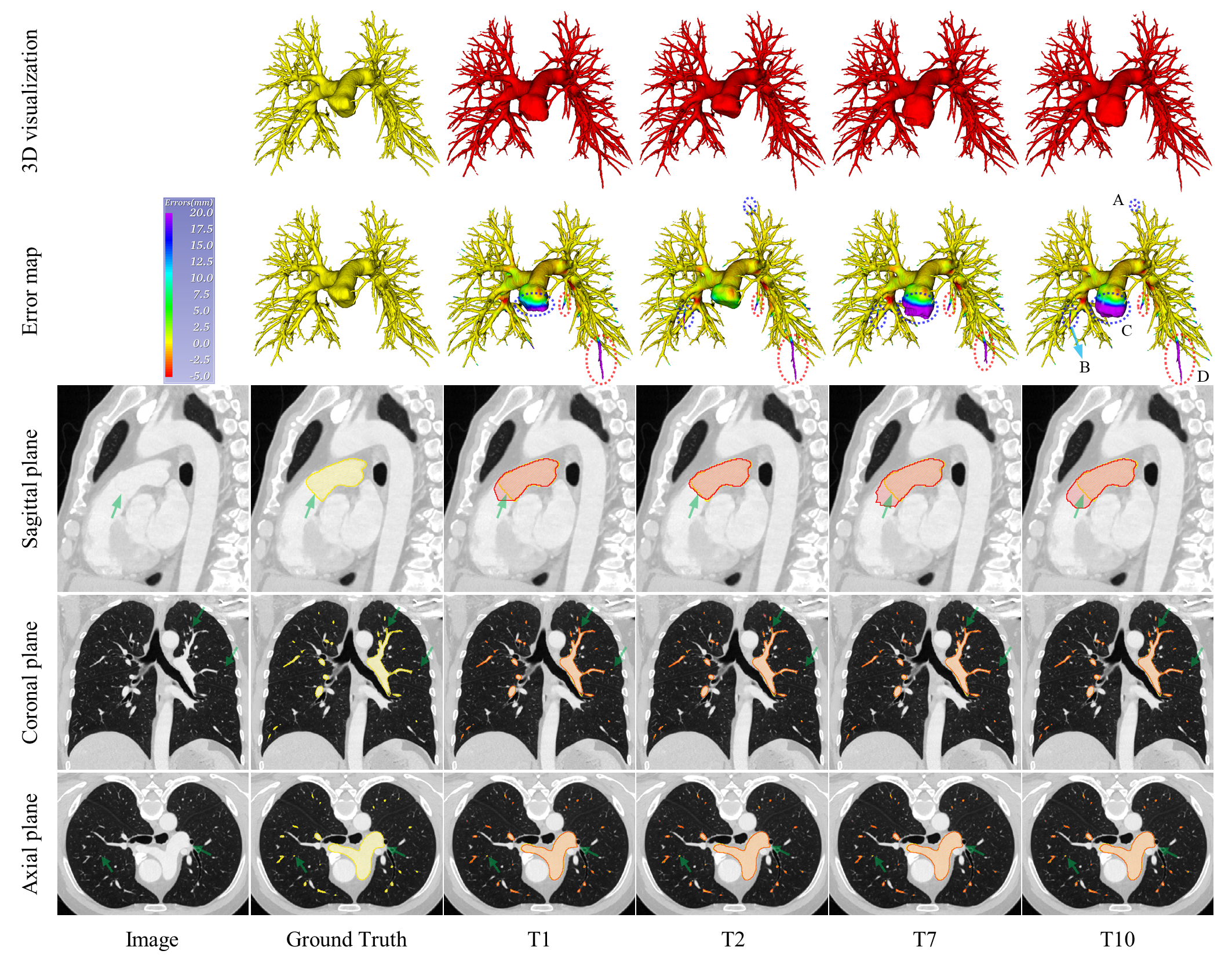}
  \caption[]{3D visualization of segmentation results (the first row), error maps between predicted results and ground truth (the second row), and three-view segmentation results (the last three rows) for representative teams. Left to right: original images, the ground truth of PA, and segmentation results from teams T1, T2, T7, and T10. The error maps represent the distance between the segmentation result and the closest position of the ground truth. The blue dashed regions in the second row are prone to segmenting false positive results. And, the red dotted areas are predicted to appear in most teams. The light green arrows point to error-prone areas in the last three rows.}
  \label{visualization}
\end{figure*}

Specifically, as shown in Table \ref{tab:all_results} and Fig. \ref{HD95_blox} (b) and (e), T1 obtained the top HD95 score for branch PA and was significantly superior to most of the teams. Meanwhile, the DSC score of T7 is not low. Therefore, the segmentation error in the branch PA region of T7 (Fig. \ref{visualization}) is relatively less. In contrast, both the DSC and the HD95 scores of T10 are not very good (Table \ref{tab:all_results}). Therefore, T10 has more segmentation errors in the main and branch PA area (Fig. \ref{visualization}).

As shown in the statistical significance maps (Fig. \ref{DSC_blox} (d - f) and Fig. \ref{HD95_blox} (d - f)), only the top 3 teams achieved significantly better DSC and HD95 scores for main PA, while most of the teams obtained those for branch PA. These results highlight that it is a more robust approach to evaluate branch PA and it is reasonable to assign more weight to branch PA.

\subsubsection{Running time and max used GPU memory analysis}
\begin{figure*}
  \centering
  \includegraphics[height=7.3cm,width=8cm]{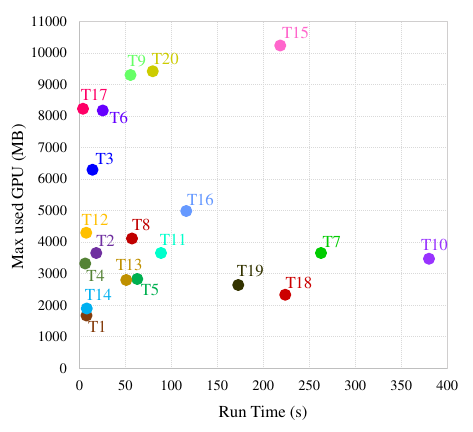}
  \caption[]{Comparative analysis of running time and max used GPU metrics of the top 20 teams.}
  \label{RTandGPU}
\end{figure*}

RT and GPU are efficiency-oriented metrics that are very relevant for clinical applications. Both the RT and GPU measure scores of the top 20 teams were shown in Fig. \ref{RTandGPU} for better presentation. Some teams, such as T3, T6 and T17, paid more attention to the RT metric and ignored the GPU metric. On the contrary, some teams, such as T19, T18, T7 and T10, paid more attention to the GPU metric and ignored the RT metric. Importantly, some teams, such as T1 and T14, could achieve superior RT and GPU scores simultaneously. As a result, if properly handled, the inference speed for PA segmentation can be accelerated with reduced GPU usage in clinical applications.

\section{Discussion}
\subsection{Ranking stability analysis}
\begin{figure*}
  \centering
  \includegraphics[width=16cm]{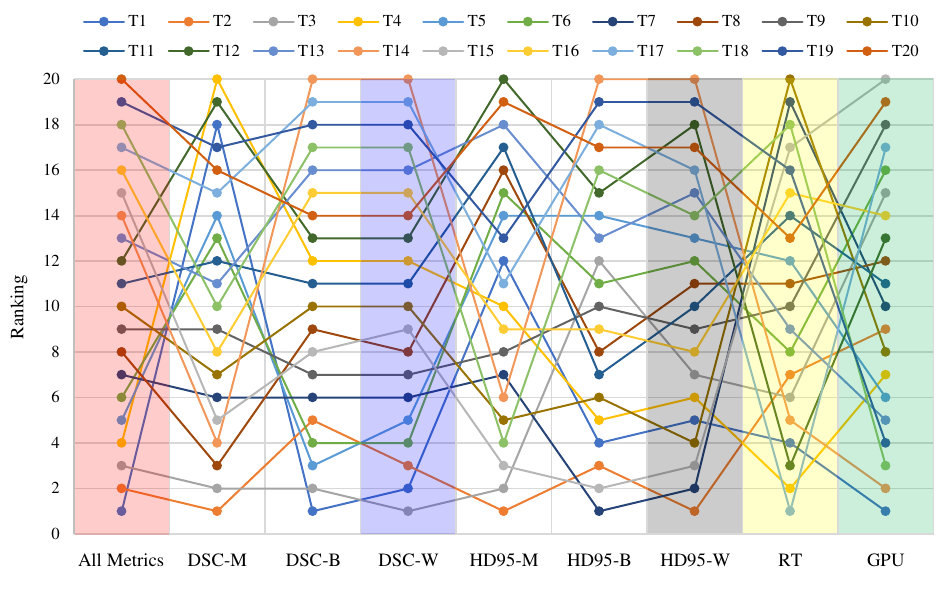}
  \caption[]{Ranking stability analysis for the top 20 teams with individual and ensemble metrics. According to the official ranking scheme, "All Metrics" (with red shading) is the ensemble of four metrics (DSC for weighted PA, HD95 for weighted PA, RT and GPU Memory with blue, grey, yellow and green shading, respectively). M, B and W denote the main, branch and weighted PA, respectively.}
  \label{ranking20WithBranch}
\end{figure*}

For a better balance between segmentation accuracy and efficiency, the four complementary metrics (DSC, HD95, RT and GPU Memory) were given the same weight in the final ranking scheme. Furtherly, to focus on segmentation accuracy for branch PA, more weight was assigned to branch PA and the weights were 80\% and 20\% for branch and main PA, respectively. In order to examine how individual metrics and ROIs (main and branch PA) affect the final rankings, the ranking stability of the top 20 teams is shown in Fig. \ref{ranking20WithBranch}. The results have large fluctuations for metrics (with shade) and ROIs (without shade). This is because different teams have different priorities. For example, T17 focused on the RT metric and obtained first place on RT, while the rankings of other metrics were not so favorable. Additionally, T1 concentrated on branch PA segmentation and ranked first on the DSC-B, while only ranking 18th on the DSC-M and 2nd on the DSC-W. In contrast, T3 gave the same priorities to main and branch PA and ranked 2nd on both DSC-M and DSC-B, while ranking 1st on DSC-W. Therefore, a more comprehensive consideration of various metrics and ROI will lead to better performance. This is also an appropriate direction for future algorithm development.

% \begin{figure*}
%   \centering
%   \includegraphics[width=8cm]{images/Figure5_ranking stability analysis.pdf}
%   \caption[]{Ranking stability analysis for the top 10 teams with different metrics. The all metrics denote the ensemble of the four metrics by the official ranking scheme, and RT and GPU denote the running time and max used GPU memory, respectively.}
%   \label{fig5:ranking}
% \end{figure*}

% \begin{figure*}
%   \centering
%   \includegraphics[width=8cm]{images/Figure5_ranking stability analysis top20.pdf}
%   \caption[]{Ranking stability analysis for the top 20 teams with different metrics. The all metrics denote the ensemble of the four metrics by the official ranking scheme, and RT and GPU denote the running time and max used GPU memory, respectively.}
%   \label{fig5:ranking20}
% \end{figure*}

\begin{figure*}
  \centering
  \includegraphics[width=16cm]{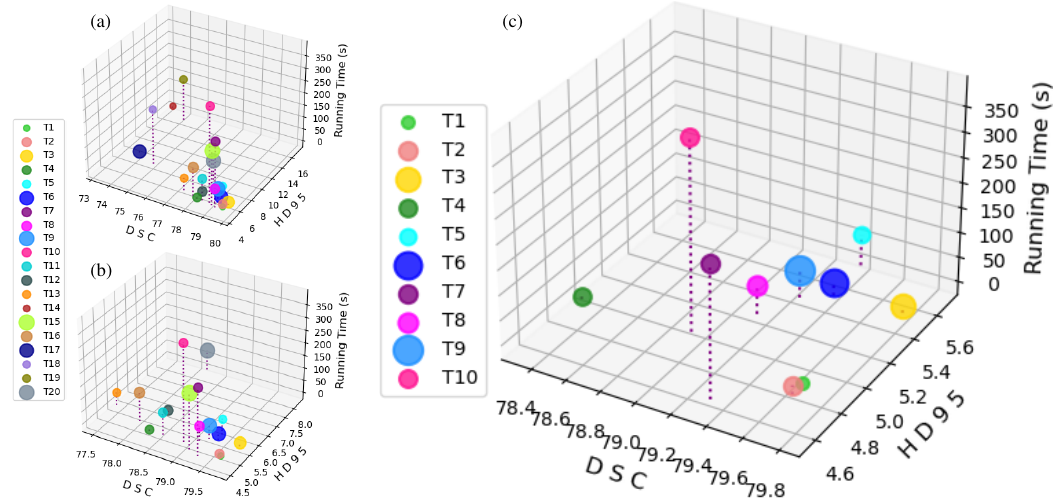}
  \caption[]{3D bubble plots of the scores, including the average DSC (x-axis) and HD95 (y-axis), Running Time (z-axis) and GPU consumption (various sphere sizes), of the top 20 teams (a), top 20 teams excluding the four teams with much lower DSC and HD95 scores (b) and top 10 teams (c). The sphere size is proportional to GPU consumption. Different colors indicate different teams. As a result of their larger DSC, less HD95, lower Running Time, and smaller sphere, these teams achieve a good balance between accuracy and efficiency.}
  \label{3Dresults}
\end{figure*}

\subsection{Balance between segmentation accuracy and efficiency}
The scores of the four metrics for the top 20 teams are shown in 3D bubble plots in Fig. \ref{3Dresults}. The teams with larger DSC, less HD95, lower RT, and smaller sphere in Fig. \ref{3Dresults}, such as T1 and T2, can achieve a suitable balance between accuracy and efficiency. However, some teams, such as T14, T17, T18 and T19, prefer to optimize efficiency (better GPU or RT performance) at the expense of accuracy (pretty poor DSC and HD95 performance). Some teams, such as T15, tend to optimize accuracy (high DSC and HD95 scores) at the expense of efficiency (poor GPU and RT scores). Another example, although T7 achieved the highest HD95 score and good DSC performance, it ignored efficiency.

\begin{table*}[]
  \centering
  \caption{Key techniques of the top 10 teams. Abbreviations: a) Preprocessing: Intensity Clipping (IC), Normalization (N), Cropping (C), Resampling (RS), Lung area extraction (LAE); b) Data augmentation: Rotation (R), Flipping (F), Scaling (S), Deformation (D), Intensity (I), Mirror (M), Random cropping (RC), Random noise (RN); c) Network design: Coarse-to-fine Framework (CF), Residual Block (RB), Deep supervision (DS), Auxiliary Loss (AL); d) Inference: LAE, Test-time augmentation (TTA), Multiple models ensemble (ME); e) Postprocessing: Connected Component Analysis (CCA), Lung area to reduce false positives (RFP), Morphological closure (MC).}
  \label{tab:method}
  \setlength{\tabcolsep}{1.5mm}
  \begin{tabular}{ccccccccccccccccccccccccc}
    \hline
    \toprule
    \multicolumn{1}{c}{\multirow{2}{*}{Method}} & \multicolumn{5}{c}{Preprocessing} & \multicolumn{8}{c}{Data augmentation} & \multicolumn{4}{c}{Network design} & \multicolumn{3}{c}{Inference} & \multicolumn{3}{c}{Postprocessing} \\ 
    \cmidrule(r){2-6}
    \cmidrule(r){7-14} 
    \cmidrule(r){15-18}
    \cmidrule(r){19-21}
    \cmidrule(r){22-24}
    \multicolumn{1}{c}{} & IC & N & C & RS & \multicolumn{1}{c}{LC} & R & F & S & D & I & M & RC & \multicolumn{1}{c}{RN} & CF & RB & DS & \multicolumn{1}{c}{AL} & LC & TTA & \multicolumn{1}{c}{ME} & CCA  & RFP & MC \\ \hline 
    T1 &$\checkmark$&$\checkmark$&$\checkmark$&  &$\checkmark$&$\checkmark$&  &  &$\checkmark$&$\checkmark$&$\checkmark$&$\checkmark$&  &  &$\checkmark$&$\checkmark$&$\checkmark$&$\checkmark$&  &$\checkmark$&$\checkmark$&$\checkmark$&  \\
    T2 &$\checkmark$&$\checkmark$&$\checkmark$&$\checkmark$&$\checkmark$&$\checkmark$&$\checkmark$&$\checkmark$&$\checkmark$&$\checkmark$&$\checkmark$&  &$\checkmark$&  &  &$\checkmark$&  &$\checkmark$&$\checkmark$&  &  &  &  \\
    T3 &$\checkmark$&$\checkmark$&$\checkmark$&$\checkmark$&$\checkmark$&$\checkmark$&$\checkmark$&$\checkmark$&$\checkmark$&$\checkmark$&  &$\checkmark$&$\checkmark$&  &$\checkmark$&$\checkmark$&  &$\checkmark$&$\checkmark$&  &  &  &  \\
    T4 &  &$\checkmark$&$\checkmark$&  &$\checkmark$&  &$\checkmark$&  &  &  &  &$\checkmark$&  &  &$\checkmark$&  &$\checkmark$&$\checkmark$&$\checkmark$&  &  &$\checkmark$&  \\
    T5 &$\checkmark$&$\checkmark$&$\checkmark$&$\checkmark$&  &$\checkmark$&  &  &$\checkmark$&$\checkmark$&  &  &  &  &  &$\checkmark$&  &  &  &  &  &  &  \\
    T6 &$\checkmark$&$\checkmark$&$\checkmark$&  &$\checkmark$&$\checkmark$&  &$\checkmark$&  &$\checkmark$&$\checkmark$&  &  &  &  &$\checkmark$&  &$\checkmark$&  &$\checkmark$&  &$\checkmark$&  \\
    T7 &$\checkmark$&$\checkmark$&$\checkmark$&$\checkmark$&  &$\checkmark$&  &$\checkmark$&  &$\checkmark$&$\checkmark$&  &$\checkmark$&$\checkmark$&  &$\checkmark$&  &  &  &$\checkmark$&$\checkmark$&  &$\checkmark$\\
    T8 &  &$\checkmark$&  &$\checkmark$&  &$\checkmark$&$\checkmark$&$\checkmark$&  &$\checkmark$&$\checkmark$&  &  &  &  &$\checkmark$&$\checkmark$&  &  &  &  &  &  \\
    T9 &  &$\checkmark$&$\checkmark$&$\checkmark$&  &  &$\checkmark$&$\checkmark$&  &$\checkmark$&  &$\checkmark$&$\checkmark$&$\checkmark$&  &  &  &  &$\checkmark$&$\checkmark$&$\checkmark$&  &  \\
    T10 &  &$\checkmark$&$\checkmark$&$\checkmark$&  &$\checkmark$&  &$\checkmark$&$\checkmark$&  &$\checkmark$&  &  &  &  &  &  &  &  &  &$\checkmark$&  &  \\ \hline \toprule
  \end{tabular}
\end{table*}

\subsection{Strategies for improving the segmentation accuracy}
As shown in Fig. \ref{visualization}, for main PA, the error-prone regions for segmentation are at the root of the main PA. For branch PA, the error-prone regions for segmentation are in some branch PA segments and the ends of the branch PA. In addition, the PA intertwined with pulmonary veins or other vessels are also more difficult to segment (Fig. \ref{visualization}). Therefore, the complex topology of the PA and large intra- and inter-class imbalances make it difficult to segment the PA effectively and efficiently. We outline the strategies of the Top 10 teams in Table \ref{tab:method} and analyze common approaches as follows. 

\textbf{(1) Data Preprocessing}  Most teams use Intensity Clipping (IC) methods and normalize (N) the intensity to [0,1] or zero mean with unit standard deviation. This contributed to the reduction of intensity variability among the different samples. Since the in-plane resolution of the original image reaches 512 $\times$ 512, it can hardly be fed directly into the network because of the huge memory consumption. Most teams crop (C) the image by lung region to get ROI and then resample (RS) to the network input size. In particular, the top teams use lung area extraction (LAE) as part of their data preprocessing by applying the threshold method and connected region analysis. Note that T3 leveraged a pre-trained network to implement LAE and T9 adopted a coarse-to-fine strategy to make LAE attend the training of PA segmentation. LAE is beneficial for the network to focus on PA area to get better accuracy and efficiency.

\textbf{(2) Data Augmentation}  Extensive dada augmentation (e.g. rotation (R), flipping (F), scaling (S), deformation (D), intensity (I), mirror (M), random cropping (RC), random noise (RN), etc.) were used to improve segmentation accuracy by most top teams. For example, T2 experimentally confirmed that better elastic deformation and global intensity perturbation parameters for better generalization of the PARSE data.

\textbf{(3) Network Design}  All the top 10 teams use U-Net\citep{unet2015} or its improved framework, such as nnU-Net\citep{nnunet}. 
This is also the primary choice for the winning algorithms in other segmentation challenges\citep{nnunet,oreiller2022head,timmins2021comparing,heller2021state,ma2022fast}. It is worth noting that simply copying and applying the original nnU-Net still produces acceptable results in segmentation accuracy, but with lower RT scores, such as T7 and T10. This is because nnU-Net uses many complex pre- and post-processing techniques\citep{nnunet}, which can consume a lot of time. In addition, T2 modified nnU-Net\citep{nnunet} by data augmentation, over-sampling mechanism and reducing TTA views, resulting in optimal segmentation accuracy and efficiency. Besides, T1 designed a lightweight U-Net framework with residual blocks and also achieved superior segmentation accuracy and efficiency. Furthermore, deep supervision (DS) is a commonly used technique for top teams to improve the accuracy of PA segmentation.

\textbf{(4) Coarse-to-fine Framework (C2F)}  C2F includes a coarse localization network and a fine segmentation network, which is a popular method in many segmentation tasks\citep{ma2022fast}. T9 explicitly uses the C2F method. Actually, most teams implicitly used the C2F method by LAE during the data preprocessing phase, especially for the top 5 teams.

\textbf{(5) Auxiliary Loss (AL)}  In addition to the commonly used Dice loss or CE loss, some teams use additional losses to aid in training. For example, T1 modified General Union Loss\citep{zheng2021alleviating} and made it train with weighted CE loss. In another example, T4 uses Focal Loss, BCE Loss and L1 Loss.

\textbf{(6) Inference}  During the inference phase, LAE was still an effective technique to speed up the prediction and the top 4 teams used LAE. In nnU-Net\citep{nnunet}, test-time augmentation (TTA) is applied by mirroring along all axes, and has been proven to enhance segmentation accuracy. However, TTA is detrimental to segmentation efficiency. T2 evaluated the segmentation accuracy of each augmented view (i.e. flipping the explicit axis) and ensembled the four most accurate views to establish a trade-off between segmentation accuracy and efficiency. Model ensemble (ME) is a frequently used method in the medical image segmentation challenge. For example, all the winning algorithms leveraged ME method in the 10 MICCAI 2020 segmentation challenges\citep{ma2022fast}. Nevertheless, ME will produce large model sizes and consume computational resources, which is not appropriate for the efficiency metrics in this challenge. Therefore, the top 5 teams did not use the ME method. It is worth noting that T1 introduced "parameter averaging" from multiple models instead of naive ME, which wouldn't increase memory consumption and computational burden during inference.

\textbf{(7) Postprocessing}  During postprocessing, connected component analysis (CCA) and LAE were performed to reduce false positives (RFP) prediction. The former was used by T1, T7, T9 and T10, while the latter was used by T1, T4 and T6. In addition, T7 also leveraged the morphological closure (MC) operation to fill the hole in the LA to reduce false negative predictions.

\subsubsection{Strategies for improving the segmentation accuracy for branch PA}
To improve the segmentation accuracy of branch PA, some teams (such as T1 and T2) offered more sampling probabilities for branch PA to make the network pay more attention to the branch. 

\subsection{Strategies for improving the segmentation efficiency}
Like FLARE challenge\citep{ma2022fast}, GPU memory and RT were selected as two efficiency-oriented metrics, since the former is an extremely valuable hardware index and both are of high importance for clinical applications. To save GPU memory consumption and achieve fast inference speed, many teams optimized their algorithms as shown in Table \ref{tab:method}. We also outline the approaches of the top 10 teams to improve segmentation efficiency.

\textbf{(1) Data Preprocessing}  Extracting the lungs region from raw images and cropping them to get ROI before sending them to the network were the strategies of the top 4 teams. This would help increase accuracy and efficiency.

\textbf{(2) Network Architecture}  The lightweight design of the network is more friendly for both GPU memory consumption and RT metrics. The success of T1 which designed a light U-Net with residual blocks indicated that if trained well, a light network could have a favorable trade-off between segmentation accuracy and efficiency.

\textbf{(3) Implementation Technique}  
TTA is usually beneficial for segmentation accuracy, but it will prolong the inference time. Therefore, T2 reduced the number of views of TTA and T5 abandoned the use of TTA. Additionally, to improve inference efficiency, T3 used asynchronous threads to read and write along with main processing. 
Furthermore, T1 converted the model and data to half-precision during inference, which reduced GPU memory consumption with almost no loss of accuracy.
In addition, most top teams used patch-based inputs that sampled 3D patches from the whole ROI by sliding window with fixed voxels. Compared with the whole-volume-based strategy, patch-based inputs will take more time to predict the final results, as suggested by FLARE challenge\citep{ma2022fast}. Thus, whole-volume-based strategy is worth considering in future algorithms.

\subsection{Limitations and future work}
As the challenge proceeded, it was found that there was some label noise in the main PA in the training set. However, the data and labels are the same for all challenge participants. Moreover, we corrected the relevant noise in the test set and the test set was more accurate. Therefore, it was fair for the participants.
Additionally, despite our best efforts, accurate labeling of the branch PA (especially their end part) is difficult, and thus some errors may appear on the labels. An example is shown in the red dashed area in Fig. \ref{visualization}, which could be segmented by most teams. Upon manual inspection, these regions are indeed PA locations. Consequently, the deep network may segment more PA. In future research, more branch PA ends can be labeled and labeled more accurately by combining traditional segmentation methods with the deep network. Therefore, labeling branch PA is an urgent issue that needs to be addressed in the future.

In this challenge dataset, only one disease (i.e., pulmonary nodules) was included. In future versions of the PARSE challenge, we will expand the size of the dataset and add more types of disease as well as the healthy population. Additionally, in this dataset, only segmentation of the pulmonary artery was taken into account. In subsequent challenges, we will also consider segmentation of the pulmonary veins and even the pulmonary airways. Furthermore, for this challenge dataset, only two levels of pulmonary arteries were analyzed. In future challenges, more levels could be considered, as in the literature\citep{poletti2022automated}.

\section{Conclusion}
In summary, we have organized the first international public PA segmentation challenge, which focuses on both the main and branch PA segmentation for high accuracy and efficiency. The results demonstrate the promise of data preprocessing, data augmentation, network architecture and additional loss function design, inference and data postprocessing techniques for accurate PA segmentation with less inference time and lower GPU memory consumption. Furthermore, we note that the segmentation accuracy of PA needs to be improved, especially for branch PA. In addition, a higher metric score (e.g., inference running time) does not mean that other metric scores (e.g., DSC and GPU memory consumption) are also better. Therefore, a better trade-off between segmentation accuracy and efficiency is needed in future research.

%%Harvard
\bibliographystyle{elsarticle-harv}\biboptions{authoryear}
\bibliography{medima-template.bib}

\end{document}